# A density-wave-like transition in the polycrystalline $V_3Sb_2$ sample with bilayer kagome lattice

N. N. Wang,[1,2] Y. H. Gu,[1,2] M. A. McGuire,[3] J. Q. Yan,[3] L. F. Shi,[1,2] Q. Cui,[1,2] K. Y. Chen,[1,2] Y. X. Wang,[1,2] H. Zhang,[1,2] H. X. Yang,[1,2] X. L. Dong,[1,2] K. Jiang,[1,2] J. P. Hu,[1,2] B. S. Wang,[1,2] J. P. Sun[1,2] and J.-G. Cheng[1,2*]

[1]*Beijing National Laboratory for Condensed Matter Physics and Institute of Physics, Chinese Academy of Sciences, Beijing 100190, China*

[2]*School of Physical Sciences, University of Chinese Academy of Sciences, Beijing 100190, China*

[3]*Materials Science and Technology Division, Oak Ridge National Laboratory, Oak Ridge, Tennessee 37831, USA*

*Email: jgcheng@iphy.ac.cn

## Abstract

Recently, transition-metal-based kagome metals have aroused much research interest as a novel platform to explore exotic topological quantum phenomena. Here we report on the synthesis, structure, and physical properties of a bilayer kagome lattice compound $V_3Sb_2$. The polycrystalline $V_3Sb_2$ samples were synthesized by conventional solid-state-reaction method in a sealed quartz tube at temperatures below 850 °C. Measurements of magnetic susceptibility and resistivity revealed consistently a density-wave-like transition at $T_{dw} \approx 160$ K with a large thermal hysteresis, even though some sample-dependent behaviors are observed presumably due to the different preparation conditions. Upon cooling through $T_{dw}$, no strong anomaly in lattice parameters and no indication of symmetry lowering were detected in powder x-ray diffraction measurements. This transition can be suppressed completely by applying hydrostatic pressures of about 1.8 GPa, around which no sign of superconductivity is observed down to 1.5 K. Specific-heat measurements reveal a relatively large Sommerfeld coefficient $\gamma = 18.5$ mJ/mol-K$^2$, confirming the metallic ground state with moderate electronic correlations. Density functional theory calculations indicate that $V_3Sb_2$ shows a non-trivial topological crystalline property. Thus, our study makes $V_3Sb_2$ a new candidate of metallic kagome compound to study the interplay between density-wave-order, nontrivial band topology, and possible superconductivity.

## Introduction

The kagome lattice compounds consisting of corner-shared triangular network have been the focus of extensive investigations over last decades because they can host fascinating physical phenomena ranging from geometrically frustrated magnetism to nontrivial band topology.[1-13] When the localized moments are situated on the kagome lattice, frustrated spin interactions can lead to exotic magnetic ground states such as the quantum spin liquid state extensively studied in the herbertsmithite $ZnCu_3(OH)_6Cl_2$. [14-19] On the other hand, when the itinerant electrons are present, the metallic kagome lattice compounds hold the gene to achieve nontrivial electronic band structure containing Dirac cones, flat band, and van Hove singularities.[20-24] Upon proper electron filling, these features can promote novel correlated and/or topological states, such as bond density wave order,[23,25] valence-bond state,[26] chiral spin-density-wave (SDW) order,[21] nontrivial charge-density-wave (CDW) order [1,27-29], and exotic superconductivity [12,23,30], etc. Thus, much experimental effort has been devoted to the metallic 3d transition-metal-based kagome systems.

For the correlated magnetic kagome metals, the coexistence of spin and charge degrees of freedom lead to the emergence of many novel quantum phenomena with nontrivial electronic behaviors. For example, large anomalous Hall effect (AHE) and Dirac fermions have been realized in the ferromagnetic $Fe_3Sn_2$ with flat-band dominated electronic band structure [7,31,32] and the quantum-limit Chern ferromagnet $TbMn_6Sn_6$.[13] Large intrinsic AHE has been observed in ferromagnetic Weyl semimetal $Co_3Sn_2S_2$ [6,33,34] and noncollinear antiferromagnetic $Mn_3Sn$ [35]. Dirac fermions has also been revealed in the antiferromagnetic FeSn [10,36]. For the recently discovered $AV_3Sb_5$ ($A$ = K, Rb, Cs) family with quasi-2D ideal kagome layers of V ions, no static magnetic order was observed but they exhibit the coexistence and complex competition between superconductivity and chiral CDW in addition to giant AHE and pair density wave state [12,27-29,37-54]. For this latter class of materials, there are still many open issues, such as the nature of superconductivity and the mechanism of chiral CDW order.

Since the materials' realizations are scarce, it is indispensable to find more metallic kagome lattice compounds. To this end, we focus our attention on a simple binary compound $V_3Sb_2$, which is isostructural to $Fe_3Sn_2$ and contains bilayer kagome lattice of V ions [31]. This material has been known to exist at temperatures below 875±25 °C in the V-Sb phase diagram [55], but its physical properties have not been studied in detail to the best of our knowledge. In this work, we successfully synthesized nearly pure $V_3Sb_2$ polycrystalline sample and then characterized its structure, electrical transport, magnetic, and thermodynamics properties at ambient or high pressures. Our results reveal a density-wave-like transition at $T_{dw} \approx 160$ K, which is evidenced by a clear anomaly in both magnetic susceptibility and resistivity. This compound shows no long-range magnetic order and superconductivity down to 1.5 K even after the density-wave-like transition being suppressed completely by hydrostatic pressure. Since our density-functional-theory (DFT) calculations show that $V_3Sb_2$ has a non-trivial topological crystalline property, the present study indicates that it may serve as a new metallic Kagome system to study the interplay between density-wave-order, nontrivial band topology and possible superconductivity.

**Experimental and calculation details**

Polycrystalline samples of $V_3Sb_2$ were synthesized by the traditional solid-state-reaction method. The powders of V (99.99 %) and Sb (99.99 %) in the molar ratio 3:2 were thoroughly mixed and pressed into a pellet, which was then placed into a quartz tube and sealed under high vacuum (~5 ×$10^{-4}$ Pa). The sealed ampoule was slowly heated to 700 °C and held for 48 hours, and then sintered again at 750-850 °C for 48 hours several times with intermediate grinding and pelletizing. According to the V-Sb phase diagram, the LT-phase of $V_3Sb_2$ with the $Fe_3Sn_2$-type structure is stable below the peritectoid temperature of 875±25 °C, above which the V-deficient $V_{2-x}Sb$ phase is more stable [55]. Thus, we have kept the highest sintering temperatures below 850 °C in order to obtain the LT-$V_3Sb_2$ with the $Fe_3Sn_2$-type structure in the present study. Phase purity of the obtained $V_3Sb_2$ polycrystalline samples was examined by powder X-ray diffraction (XRD) at room temperature with a Cu-$K_{\alpha 1}$ radiation. In order to extract structural parameters, the refinements of the crystal structure were performed by the Rietveld method, using the FULLPROF program. Low-temperature XRD data in the temperature range 25–300 K were collected using an Oxford Phenix cryostat and a PANalytical X'Pert Pro MPD diffractometer (Cu-$K_{\alpha 1}$ radiation). Powder was mounted on an aluminum sample holder using Apiezon N-grease. Highscore Plus was used for Rietveld analysis of the XRD patterns from $V_3Sb_2$.

The magnetic properties were measured with a Quantum Design Magnetic Property Measurement System (MPMS-III). Heat capacity and electrical transport measurements were carried out by using a Quantum Design Physical Property Measurement System (PPMS, 14 T). High-pressure resistivity was measured by using a self-clamped piston-cylinder cell under various hydrostatic pressures up to 1.81 GPa [56]. Daphne 7373 was used as the pressure transmitting medium and the pressure values were determined from the relative shift of the superconducting transition temperature of Pb.

Our DFT calculations employ the Vienna ab initio simulation package (VASP) code [57] with the projector augmented wave (PAW) method.[58] The Perdew-Burke-Ernzerhof (PBE) [59] exchange-correlation functional is used in our calculations. The kinetic energy cutoff is set to be 600 eV for expanding the wave functions into a plane-wave basis in VASP calculations while the energy convergence criterion is $10^{-6}$ eV. The Γ-centered k-mesh is 8×8×8. The spin-orbital coupling is included in our DFT calculations. The calculation of topological indices employs vasp2trace program on Bilbao Crystallographic Server. [60-62]

## Results and Discussion

We have attempted to prepare pure $V_3Sb_2$ samples by varying the final sintering temperatures between 750 and 850 °C. After tries and errors, we eventually obtained two $V_3Sb_2$ polycrystalline samples with relatively high purity as examined by XRD. Hereafter, these two samples sintered at 780 °C and 830 °C are labeled as S1 and S2, respectively. Since their physical properties show some different behaviors, in the following we present the experimental results of both samples comparatively so as to underline the fact that the physical properties of $V_3Sb_2$ are sensitive to the preparation conditions.

Figure 1 shows the XRD pattern of these two $V_3Sb_2$ samples after Rietveld refinement. It confirms that the obtained samples are nearly single phase with a small amount of impurity, which can be identified as $V_3Sb$ for S1. The impurity phase is too weak to be identified for S2. The main phase of the XRD pattern can be refined by considering the hexagonal $Fe_3Sn_2$-type structure model (space group *R-3m*, No. 166) with one V position at the 18h ($x$, $y$, $z$) and two Sb positions at 6c (0, 0, $z$), respectively. As illustrated in Fig.1, the refinements are converged well with reliable factors $R_p$ = 2.66%, $R_{exp}$ = 2.48%, and $\chi^2$ = 2.28 for S1, and $R_p$ = 2.98%, $R_{exp}$ = 2.26%, and $\chi^2$ = 3.78 for S2. According to the scaling factors, the amount of $V_3Sb$ impurity in S1 was estimated to be ~1.2 wt%, which is consistent with the observed weak main peak of $V_3Sb$ as shown by asterisk in Fig. 1(a). The obtained structural parameters for both samples are listed in Table I for comparison. As can be seen, the sample S1 has a shorter $a$ but a longer $c$, leading to a slightly smaller $V$ in comparison with those of sample S2. However, the differences are quite small, with the relative difference being smaller than 0.3%. These lattice parameters are also consistent with those reported in literature for LT-$V_3Sb_2$. [55] In contrary, the positions of the atoms, especially along the $z$-axis, show a relatively large difference. Taking the V site as an example, the difference of the $z$-axis position is about 2.5% for these two samples, which result in a difference in the bond length along the $c$-axis as discussed below. It should be noted that slight variation of V/Sb ratio in samples prepared at different conditions is possible and should be responsible for the observations of sample-dependent structural and physical properties show below. In addition, synchrotron-based XRD is desirable to have more accurate determinations of the structural details.

The crystal structure of $V_3Sb_2$ is schematically shown in Fig. 2 by taking the refined structural information of S1. As seen in Fig. 2(a), the crystal structure of $V_3Sb_2$ is composed of Sb1 single layer and V-Sb2 bilayer that are stacked alternatively along the *c*-axis. The Sb1 atoms in the single layer form a 2D graphene-like honeycomb lattice, Fig. 2(b), while the V-Sb2 bilayer consists of two V kagome layers with Sb2 atoms embedded in the center of the V hexagons, Fig. 2(c). For each V kagome layer, there are two kinds of equilateral triangles with different V-V bond lengths, i.e., 2.8600(19) Å and 2.6839(19) Å for S1 [2.864(4) Å and 2.691(4) Å for S2], as shown in purple and blue in Fig. 2(d). This means that the kagome lattice is not ideal, different from that in $AV_3Sb_5$. As shown in Fig. 2(e), the V atoms of the smaller V triangles in two adjacent layers can be viewed to form a $V_6$ octahedra with an interlayer V-V distance of 2.773(4)

Å for S1 [2.870(6) Å for S2]. For S1, this value is equal to the average of two intralayer nearest-neighbor V-V bond lengths. Such a small interlayer distance should produce a relatively strong interlayer interaction. However, the interlayer V-V bond length for S2 is found to be ~ 3% longer than the average of *ab* plane, which might result in weaker interlayer interactions.

Figure 3(a, b) shows the temperature-dependent magnetic susceptibility $\chi(T)$ of these two V$_3$Sb$_2$ samples measured in the temperature range 1.8-300 K under an external magnetic field of $\mu_0H$ = 1 T. The experimental procedure is following: we first cooled the sample under zero field from room temperature to the lowest temperature, then applied 1 T field and recorded $\chi(T)$ upon heating up to room temperature, followed by measuring $\chi(T)$ upon cooling down to the lowest temperature under 1 T. They are thus denoted as the ZFCw- and FCc-$\chi(T)$, respectively. As seen in Fig. 3(a), the ZFCw- and FCc-$\chi(T)$ curves of S1 exhibit a step-like anomaly around 160 K, below which the magnetic susceptibility is reduced. In addition, an obvious thermal hysteresis is evidenced around the transition, implying a first-order nature of this transition. The transition temperatures, $T_{dw}^\chi$, defined as the maximum of d$\chi$/d$T$, are ~159 K and ~165 K for S1 in the cooling and heating processes, respectively, as shown in the top of Fig. 3(a). The $\chi(T)$ curves of S2 are basically similar to those of S1, but the transition is much wide and the thermal hysteresis becomes very weak, Fig. 3(b). In addition, the transition temperatures, $T_{dw}^\chi \approx$ 154 K and 158 K for the cooling and heating processes, are lower than those of S1. These comparisons indicate that the quality of S1 is better than that of S2. We note that the observed feature in the magnetic susceptibility is very similar to those seen in some typical CDW materials, such as 1T-TaS$_2$,[63] NbSe$_3$,[64] CuIr$_2$Te$_4$,[65] IrTe$_2$,[66] and $A$V$_3$Sb$_5$[12,37,39].

Except for the transition region, the ZFCw- and FCc-$\chi(T)$ curves are almost overlapped with each other in the whole temperature range, and show a paramagnetic behavior at high temperature and a low-temperature upturn without long-range magnetic order down to 1.8 K. As shown in Fig. 3(a), the susceptibility $\chi(T)$ of S1 in the low-temperature range 2-10 K can be fitted by a modified Curie-Weiss (CW) model, $\chi(T) = \chi_0 + C/(T - \theta_{CW})$. Here, the obtained effective moment of $\mu_{eff}$ = 0.52 $\mu_B$/V and a CW temperature of $\theta_{CW}$ = -34.3 K are similar with those of 0.22 $\mu_B$/V and -47.2 K in KV$_3$Sb$_5$.[37] Noted that the presence of impurities and/or defects should dominate the low-temperature upturn in $\chi(T)$, especially considering the polycrystalline nature of the studied sample. The $\chi^{-1}(T)$ in the high-temperature range 200-300 K, Fig. 3(c), seems to follow a linear behavior, and a CW fitting yields $\mu_{eff}$ = 3.07 $\mu_B$/V and $\theta_{CW}$ = -4820 K, respectively. Such a large $\theta_{CW}$ is not physically meaningful and is consistent with a Pauli paramagnetism.

Figure 3(d) presents the field dependence of magnetization $M(H)$ for S1 between -7 and +7 T at various temperatures. The $M(H)$ curves at all temperatures except for 1.8 K exhibit a linear dependence on the external field with no hysteresis, confirming the absence of ferromagnetic contributions to the magnetism. The magnetic moment of ~ 0.003 $\mu_B$/V at 7 T is small. These results make it significantly different from Fe$_3$Sn$_2$, but similar with the $A$V$_3$Sb$_5$ family where no magnetic order or even local moment of V ion was found. In the isostructural compound $Fe_3Sn_2$, the Fe atoms in the kagome plane exhibit strong frustrated magnetic interactions, undergoing complex magnetic phase transitions upon cooling down from a paramagnetic to a collinear ferromagnetic state at 640 K, then to a non-collinear ferromagnetic state at 350 K, and finally to a re-entrant spin glass phase at 70 K. [67] In addition, we find that the magnitudes of $\chi(T)$ and $M(H)$ for $V_3Sb_2$ are very close to those of $KV_3Sb_5$,[37] further indicating the similarity of magnetic states in these two V-based kagome systems.

The presence of a density-wave-like transition in $V_3Sb_2$ is further confirmed by the resistivity measurement. Figure 4(a) shows the temperature-dependent $\rho(T)$ of S1 measured in a thermal cycle under 0 T. As can be seen, it displays a metallic behavior in the whole temperature range and exhibits a clear hump-like anomaly around $T_{dw}^{\rho} \approx$ 157 K and 165 K in the cooling and heating processes. These transition temperatures defined from the dip of $d\rho/dT$ curve are in well agreement with those determined from the $\chi(T)$ data shown in Fig. 3(a). In addition, a similar thermal hysteresis is evidenced around $T_{dw}^{\rho}$ in $\rho(T)$. We find that the application of 10 T external magnetic field has a negligible influence on the $\rho(T)$ near $T_{dw}^{\rho}$ (data not shown). This result indicates that the transition is most likely due to the formation of CDW rather than SDW. A small positive magnetoresistance (MR) was evidenced at low temperatures. The inset of Fig. 4(a) displays the MR $\equiv \Delta\rho(H)/\rho(0) \times 100\%$ at 2 K in the field range from -14 T to 14 T. As can be seen, the MR is only ~1.2% at 2 K and 14 T, but it is non-saturating and can be well described by the expression MR $\propto H^n$ with $n = 1.35$.

Interestingly, the $\rho(T)$ of S2 is quite different with respect to that of S1 as seen in Fig. 4(b). Nonetheless, it also exhibits a pronounced anomaly centered around 150-160 K, and the transition temperatures determined from the minimum of $d\rho/dT$ are 152 K and 160 K in the cooling and heating processes, respectively. These temperatures are also consistent with those determined from the $\chi(T)$ data shown in Fig. 3(b). In comparison with S1, the transition is much stronger and extended over a wider temperature range. In addition, the resistivity below $T_{dw}$ becomes higher than that above the transition and displays a clear upturn at low temperatures. These comparisons highlight a better quality of sample S1 than that of S2, and the observed differences might be attributed to the fact that the final sintering temperature of 830 °C for S2 is closer to the peritectoid temperature.[55] In the following we thus focus our attention on sample S1. Despite of these differences, the occurrence of density-wave-like transition, most likely a CDW one, around $T_{dw} \approx 160$ K in $V_3Sb_2$ should be an intrinsic and bulk behavior.

We have performed variable-temperature powder XRD on S1 from room temperature down to 25 K in order to check if the CDW order is accompanied by obvious structural transition. Within the resolution of our instrument, no obvious peak splitting or satellite peaks are observed in the XRD patterns over the whole temperature range, as illustrated in Fig. 5(a) for a portion of the XRD patterns measured at 25 and 300 K. This indicates that the structural modifications, if exists around $T_{dw}$, should be too weak to be detected by our lab XRD, and may await for verification with high-resolution synchrotron XRD or transmission electron microscope at low temperatures. Interestingly, we find that the lattice parameter $c(T)$ exhibits a negative thermal expansion over the whole temperature range, while both $a(T)$ and $V(T)$ display normal contraction upon cooling down as shown in Fig. 5 (b). The smooth evolution of lattice parameters across $T_{dw}$ is also consistent with the absence of structural transition in $V_3Sb_2$.

To further characterize the paramagnetic and metallic ground state of $V_3Sb_2$ with a possible CDW-like transition, we performed specific-heat measurements on S1 in the wide temperature range. Figure 6 displays the $C/T$ vs $T$ of S1 from 2 to 250 K under zero field. There is no obvious specific-heat anomaly near the CDW-like transition, suggesting that the thermodynamic signature of this transition is too weak to be observed. We attributed the absence of specific-heat anomaly around $T_{dw}$ in $V_3Sb_2$ to the polycrystalline nature of the studied sample. In comparison with the single crystal, the relatively poor crystallization and the presence of grain boundaries in the polycrystalline samples would diminish considerably the specific-heat anomaly around a phase transition. This is well demonstrated in the $A$V$_3$Sb$_5$ [12,37-39]: the specific-heat anomaly around the first-order transition can be barely observed in the polycrystalline KV$_3$Sb$_5$ [37], while it becomes clearer in the single-crystal KV$_3$Sb$_5$ [38] and is much stronger in the CsV$_3$Sb$_5$ crystal [12] with a better quality than KV$_3$Sb$_5$. We also noticed that the density-wave-like transition in resistivity and susceptibility of our polycrystalline V$_3$Sb$_2$ sample is relatively broad which can further obscure the thermodynamic signature. Inset of Fig. 6 shows the plot of $C/T$ vs $T^2$ at the low-temperature range, and a linear fit to $C(T) = \gamma T + \beta T^3$ considering the electronic and lattice contributions yields the Sommerfeld coefficient $\gamma = 18.5(1)$ mJ mol$^{-1}$ K$^{-2}$ and $\beta = 0.73(1)$ mJ mol$^{-1}$ K$^{-4}$. The Debye temperature $\Theta_D \approx 237$ K can be calculated according to the relation $\Theta_D = (12\pi^4 nR/5\beta)^{1/3}$, where $R = 8.314$ J mol$^{-1}$ K$^{-1}$ is the ideal gas constant and $n = 5$ is the number of atoms per formula unit. The obtained $\gamma$ is relatively large compared with elemental metal and is close to that of KV$_3$Sb$_5$, [37] implying the moderate enhancement of effective electron mass and the presence of electronic correlations.

From the above characterizations, we can conclude that the bilayer kagome metal V$_3$Sb$_2$ is a paramagnetic metal and undergoes a CDW-like transition at $T_{dw} \approx 160$ K, which is similar to the family of $A$V$_3$Sb$_5$. But no superconductivity was observed in V$_3$Sb$_2$ at ambient pressure down to 1.8 K. To explore whether superconductivity emerges after suppressing the CDW under pressure, we measured the $\rho(T)$ of S1 under various hydrostatic pressures up to 1.8 GPa by using a piston-cylinder cell. The $\rho(T)$ and its derivative d$\rho$/d$T$ at different pressures are shown in Fig. 7(a). The evolution of the CDW-like transition with pressure can be tracked clearly from the resistivity anomaly. With increasing pressure gradually, the anomaly in $\rho(T)$ and the corresponding $T_{dw}$ determined from the minimum of d$\rho$/d$T$ continuously move to lower temperatures. The pressure dependence of the determined $T_{dw}$ is plotted in Fig. 7(b). As can be seen, $T_{dw}$ is suppressed to about 60 K at 1.6 GPa, above which it cannot be clearly distinguished in both $\rho(T)$ and d$\rho$/d$T$, implying a complete suppression of CDW-like order above 1.6 GPa. However, no sign of superconductivity can be observed down to 1.5 K accompanying the complete suppression of CDW-like order. Whether superconductivity can be realized at much lower temperatures or on high-quality single-crystal samples deserves further studies. Although the above characterizations have revealed a first-order character for the CDW-like transition at ambient pressure, the resistivity anomaly around $T_{dw}$ is weakened gradually by applying pressure, Fig. 7(a). If a crossover from first-order to second-order transition can take place, a possible putative quantum critical point can be realized under pressure. When the high-quality $V_3Sb_2$ single crystals become available, further high-pressure studies are desirable for in-depth investigations on the pressure-induced quantum critical phase transition.

Finally, we calculated the electronic structure of $V_3Sb_2$ by first-principles calculations. As shown in Fig. 8, the $V_3Sb_2$ shows a metallic nature and its Fermi surface is mainly composed by V's $d$ orbitals. In $V_3Sb_2$, there are two kinds of antimony, which are chemically inequivalent, forming different bands. The hybridization between Sb2 and V is relatively stronger than that between Sb1 and V, because Sb2 are intralayer with V atoms. Meanwhile, we calculated the topological indices of $V_3Sb_2$, which are $z_{2w,1}=0$, $z_{2w,2}=0$, $z_{2w,3}=0$ and $z_4=2$, showing a non-trivial topological crystalline property. [60-62]

The present work is a preliminary study on the physical properties of $V_3Sb_2$ polycrystalline samples and leaves many open questions for the future experimental and theoretical studies. For example, electrical transport and magnetic properties at much lower temperatures should be measured to explore possible superconductivity or magnetic order. Low-temperature synchrotron XRD, transmission electron microscope, optical spectroscopy, and scanning tunneling microscope measurements should be performed to elucidate the nature of the density-wave-like transition. Moreover, to acquire high-quality single-crystal samples is mandatory for in-depth characterizations of the intrinsic electronic structure by using the angle-resolved photoemission spectroscopy.

## Conclusion

In summary, we have synthesized the polycrystalline sample of $V_3Sb_2$ with bilayer kagome lattice of V atoms through a traditional solid-state-reaction method and characterized its structural, electrical transport, magnetic, and thermodynamic properties via x-ray powder diffraction, resistivity, magnetic susceptibility, and specific heat measurements. We observed no long-range magnetic order in $V_3Sb_2$ above 1.8 K, which is completely different from the isostructural $Fe_3Sn_2$ but similar to the kagome metal $AV_3Sb_5$ ($A$ = K, Rb, Cs) family. In addition, a density-wave-like anomaly was evidenced around 160 K in $V_3Sb_2$, making it more similar to the $AV_3Sb_5$ ($A$ = K, Rb, Cs) family. Moreover, we find that the density-wave-like transition can be gradually suppressed by pressure but no sign of superconductivity can be observed down to 1.5 K. We proposed that $V_3Sb_2$ is a novel candidate kagome metal to study the interplay between density-wave-order, nontrivial band topology and possible superconductivity.

## Note added:

During the preparation of this manuscript, we noticed that Shi *et al*. reported the synthesis and characterizations of the "V$_6$Sb$_4$" single crystal in a recent preprint arXiv: 2110.09782. [68] The "V$_6$Sb$_4$" single crystal was grown at a high temperature of 1100 °C and does not exhibit any density-wave-like transition, different from what we observed in the low-temperature phase of V$_3$Sb$_2$ in the present work. We attributed the observed different behaviors to the different sintering temperatures.

## Acknowledgements

This work is supported by National Key R&D Program of China (2018YFA0305700, 2018YFA0305800), the National Natural Science Foundation of China (Grant Nos. 12025408, 11874400, 11834016, 11921004, 11888101, 11904391), the Beijing Natural Science Foundation (Z190008), the Strategic Priority Research Program and Key Research Program of Frontier Sciences of CAS (XDB25000000, XDB33000000 and QYZDB-SSW-SLH013), and the CAS Interdisciplinary Innovation Team (JCTD-201-01). Work at Oak Ridge National Laboratory was supported by the U.S. Department of Energy, Office of Science, Basic Energy Sciences, Materials Sciences and Engineering Division.## References

[1] Guo H M and Franz M 2009 *Phys. Rev. B* **80** 113102
[2] Balents L 2010 *Nature* **464** 199
[3] Yan S, Huse D A, and White S R 2011 *Science* **332** 1173
[4] Depenbrock S, McCulloch I P, and Schollwoeck U 2012 *Phys. Rev. Lett.* **109** 067201
[5] Han T H, Helton J S, Chu S, Nocera D G, Rodriguez-Rivera J A, Broholm C, and Lee Y S 2012 *Nature* **492** 406
[6] Liu E, Sun Y, Kumar N, Muechler L, Sun A, Jiao L, Yang S Y, Liu D, Liang A, Xu Q, Kroder J, Suess V, Borrmann H, Shekhar C, Wang Z S, Xi C Y, Wang W H, Schnelle W, Wirth S, Chen Y, Goennenwein S T B, and Felser C 2018 *Nat. Phys.* **14** 1125
[7] Ye L, Kang M, Liu J W, von Cube F, Wicker C R, Suzuki T, Jozwiak C, Bostwick A, Rotenberg E, Bell D C, Fu L, Comin R, and Checkelsky J G 2018 *Nature* **555** 638
[8] Yin J X, Zhang S S, Li H, Jiang K, Chang G Q, Zhang B J, Lian B, Xiang C, Belopolski I, Zheng H, Cochran T A, Xu S Y, Bian G, Liu K, Chang T R, Lin H, Lu Z Y, Wang Z Q, Jia S, Wang W H, and Hasan M Z 2018 *Nature* **562** 91
[9] Kang M, Fang S, Ye L, Po H C, Denlinger J, Jozwiak C, Bostwick A, Rotenberg E, Kaxiras E, Checkelsky J G, and Comin R 2020 *Nat. Commun.* **11** 4004
[10] Kang M, Ye L, Fang S, You J-S, Levitan A, Han M, Facio J I, Jozwiak C, Bostwick A, Rotenberg E, Chan M K, McDonald R D, Graf D, Kaznatcheev K, Vescovo E, Bell D C, Kaxiras E, van den Brink J, Richter M, Prasad Ghimire M, Checkelsky J G, and Comin R 2020 *Nat. Mater.* **19** 163
[11] Liu Z H, Li M, Wang Q, Wang G W, Wen C H P, Jiang K, Lu X L, Yan S C, Huang Y B, Shen D W, Yin J X, Wang Z Q, Yin Z P, Lei H C, and Wang S C 2020 *Nat. Commun.* **11** 4002
[12] Ortiz B R, Teicher S M L, Hu Y, Zuo J L, Sarte P M, Schueller E C, Abeykoon A M M, Krogstad M J, Rosenkranz S, Osborn R, Seshadri R, Balents L, He J, and Wilson S D 2020 *Phys. Rev. Lett.* **125** 247002

[13] Yin J X, Ma W L, Cochran T A, Xu X T, Zhang S S, Tien H J, Shumiya N, Cheng G M, Jiang K, Lian B, Song Z, Chang G Q, Belopolski I, Multer D, Litskevich M, Cheng Z-J, Yang X P, Swidler B, Zhou H B, Lin H, Neupert T, Wang Z Q, Yao N, Chang T-R, Jia S, and Zahid Hasan M 2020 *Nature* **583** 533
[14] Shores M P, Nytko E A, Bartlett B M, and Nocera D G 2005 *J. Am. Chem. Soc.* **127** 13462
[15] Helton J S, Matan K, Shores M P, Nytko E A, Bartlett B M, Yoshida Y, Takano Y, Suslov A, Qiu Y, Chung J H, Nocera D G, and Lee Y S 2007 *Phys. Rev. Lett.* **98** 107204
[16] Ran Y, Hermele M, Lee P A, and Wen X-G 2007 *Phys. Rev. Lett.* **98** 117205
[17] Rigol M and Singh R R P 2007 *Phys. Rev. Lett.* **98** 207204
[18] Punk M, Chowdhury D, and Sachdev S 2014 *Nat. Phys.* **10** 289
[19] Kelly Z A, Gallagher M J, and McQueen T M 2016 *Phys. Rev. X* **6** 041007
[20] O'Brien A, Pollmann F, and Fulde P 2010 *Phys. Rev. B* **81** 235115
[21] Yu S L and Li J X 2012 *Phys. Rev. B* **85** 144402
[22] Kiesel M L, Platt C, and Thomale R 2013 *Phys. Rev. Lett.* **110** 126405
[23] Wang W S, Li Z Z, Xiang Y Y, and Wang Q H 2013 *Phys. Rev. B* **87** 115135
[24] Mazin I I, Jeschke H O, Lechermann F, Lee H, Fink M, Thomale R, and Valenti R 2014 *Nat. Commun.* **5** 4261
[25] Isakov S V, Wessel S, Melko R G, Sengupta K, and Kim Y B 2006 *Phys. Rev. Lett.* **97** 147202
[26] Guertler S 2014 *Phys. Rev. B* **90** 081105
[27] Li H X, Zhang T T, Yilmaz T, Pai Y Y, Marvinney C E, Said A, Yin Q W, Gong C S, Tu Z J, Vescovo E, Nelson C S, Moore R G, Murakami S, Lei H C, Lee H N, Lawrie B J, and Miao H 2021 *Phys. Rev. X* **11** 031050
[28] Wang Z X, Wu Q, Yin Q W, Gong C S, Tu Z J, Lin T, Liu Q M, Shi L Y, Zhang S J, Wu D, Lei H C, Dong T, and Wang N L 2021 *Phys. Rev. B* **104** 165110
[29] Yu F H, Ma D H, Zhuo W Z, Liu S Q, Wen X K, Lei B, Ying J J, and Chen X H 2021 *Nat. Commun.* **12** 3645
[30] Ko W-H, Lee P A, and Wen X-G 2009 *Phys. Rev. B* **79** 214502
[31] Wang Q, Sun S S, Zhang X, Pang F, and Lei H C 2016 *Phys. Rev. B* **94** 075135
[32] Lin Z Y, Choi J H, Zhang Q, Qin W, Yi S, Wang P D, Li L, Wang Y F, Zhang H, Sun Z, Wei L M, Zhang S B, Guo T F, Lu Q Y, Cho J H, Zeng C G, and Zhang Z Y 2018 *Phys. Rev. Lett.* **121** 096401
[33] Wang Q, Xu Y F, Lou R, Liu Z H, Li M, Huang Y B, Shen D W, Weng H M, Wang S C, and Lei H C 2018 *Nat. Commun.* **9** 4212
[34] Liu Z Y, Zhang T, Xu S X, Yang P T, Lei H C, Yu. S, Uwatoko Y, Wang B S, Wen H M, Sun J P, and Cheng J G 2020 *Phys. Rev. Mater.* **4** 044203
[35] Nakatsuji S, Kiyohara N, and Higo T 2015 *Nature* **527** 212
[36] Lin Z Y, Wang C Z, Wang P D, Yi S, Li L, Zhang Q, Wang Y, Wang Z Y, Huang H, Sun Y, Huang Y B, Shen D W, Feng D L, Sun Z, Cho J-H, Zeng C G, and Zhang Z Y 2020 *Phys. Rev. B* **102** 155103
[37] Ortiz B R, Gomes L C, Morey J R, Winiarski M, Bordelon M, Mangum J S, Oswald L W H, Rodriguez-Rivera J A, Neilson J R, Wilson S D, Ertekin E, McQueen T M, and Toberer E S 2019 *Phys. Rev. Mater.* **3** 094407
[38] Ortiz B R, Sarte P M, Kenney E M, Graf M J, Teicher S M L, Seshadri R, and Wilson S D

2020 *Phys. Rev. Mater.* **5** 034801

[39] Yin Q W, Tu Z J, Gong C S, Fu Y, Yan S H, and Lei H C 2021 *Chin. Phys. Lett.* **38** 037403

[40] Yang S Y, Wang Y J, Ortiz B R, Liu D, Gayles J, Derunova E, Gonzalez-Hernandez R, Smejkal L, Chen Y L, Parkin S S P, Wilson S D, Toberer E S, McQueen T, and Ali M N 2020 *Sci. Adv.* **6** eabb6003

[41] Chen H, Yang H T, Hu B, Zhao Z, Yuan J, Xing Y Q, Qian G J, Huang Z H, Li G, Ye Y H, Ma S, Ni S L, Zhang H, Yin Q W, Gong C S, Tu Z J, Lei H C, Tan H X, Zhou S, Shen C M, Dong X L, Yan B H, Wang Z Q, and Gao H J 2021 *Nature* **599** 222

[42] Chen K Y, Wang N N, Yin Q W, Gu Y H, Jiang K, Tu Z J, Gong C S, Uwatoko Y, Sun J P, Lei H C, Hu J P, and Cheng J G 2021 *Phys. Rev. Lett.* **126** 247001

[43] Wang N N, Chen K Y, Yin Q W, Ma Y N N, Pan B Y, Yang X, Ji X Y, Wu S L, Shan P F, Xu S X, Tu Z J, Gong C S, Liu G T, Li G, Uwatoko Y, Dong X L, Lei H C, Sun J P, and Cheng J G 2021 *Phys. Rev. Research* **3** 043018

[44] Du F, Luo S S, Ortiz B R, Chen Y, Duan W Y, Zhang D T, Lu X, Wilson S D, Song Y, and Yuan H Q 2021 *Phys. Rev. B* **103** L220504

[45] Jiang Y X, Yin J X, Denner M M, Shumiya N, Ortiz B R, Xu G, Guguchia Z, He J, Hossain M S, Liu X X, Ruff J, Kautzsch L, Zhang S S, Chang G, Belopolski I, Zhang Q, Cochran T A, Multer D, Litskevich M, Cheng Z J, Yang X P, Wang Z Q, Thomale R, Neupert T, Wilson S D, and Hasan M Z 2021 *Nat. Mater.* **20** 1353

[46] Kenney E M, Ortiz B R, Wang C, Wilson S D, and Graf M J 2021 *J. Phys.: Condens. Matter* **33** 235801

[47] Liang Z W, Hou X Y, Zhang F, Ma W R, Wu P, Zhang Z Y, Yu F H, Ying J J, Jiang K, Shan L, Wang Z Y, and Chen X H 2021 *Phys. Rev. X* **11** 031026

[48] Shumiya N, Hossain M S, Yin J X, Jiang Y X, Ortiz B R, Liu H X, Shi Y G, Yin Q W, Le H C, Zhan S S, Chang G Q, Zhang Q, Cochran T A, Multer D, Litskevich M, Cheng Z J, Yang X P, Guguchia Z, Wilson S D, and Hasan M Z 2021 *Phys. Rev. B* **104** 035131

[49] Wang Q, Kong P F, Shi W J, Pei C Y, Wen C H P, Gao L L, Zhao Y, Yin Q W, Wu Y S, Li G, Lei H C, Li J, Chen Y L, Yan S C, and Qi Y P 2021 *Adv. Mater.* **33** 2102813

[50] Yu F H, Wu T, Wang Z Y, Lei B, Zhuo W Z, Ying J J, and Chen X H 2021 *Phys. Rev. B* **104** L041103

[51] Zhao H, Li H, Ortiz B R, Teicher S M L, Park T, Ye M X, Wang Z Q, Balents L, Wilson S D, and Zeljkovic I 2021 *Nature* **599** 216

[52] Chen X, Zhan X H, Wang X J, Deng J, Liu X B, Chen X, Guo J G, and Chen X L 2021 *Chin. Phys. Lett.* **38** 057402

[53] Ni S L, Ma S, Zhang Y H, Yuan J, Yang H T, Lu Z Y W, Wang N N, Sun J P, Zhao Z, Li D, Liu S B, Zhang H, Chen H, Jin K, Cheng J G, Yu L, Zhou F, Dong X L, Hu J P, Gao H J, and Xian Z Z 2021 *Chin. Phys. Lett.* **38** 057403

[54] Mu C, Yin Q W, Tu Z J, Gong C S, Lei H C, Li Z, and Luo J L 2021 *Chin. Phys. Lett.* **38** 077402

[55] Failamani F, Broz P, Macciò D, Puchegger S, Müller H, Salamakha L, Michor H, Grytsiv A, Saccone A, Bauer E, Giester G, and Rogl P 2015 *Intermetallics* **65** 94

[56] Uwatoko Y, Todo S, Ueda K, Uchida A, Kosaka M, Mori N, and Matsumoto T 2002 *J. Phys.: Condens. Matter* **14** 11291

[57] Kresse G and Furthmuller J 1996 *Comp. Mater. Sci.* **6** 15

[58] Kresse G and Joubert D 1999 *Phys. Rev. B* **59** 1758
[59] Perdew J P, Burke K, and Ernzerhof M 1997 *Phys. Rev. Lett.* **78** 1396
[60] Fu L 2011 *Phys. Rev. Lett.* **106** 106802
[61] Vergniory M G, Elcoro L, Felser C, Regnault N, Bernevig B A, and Wang Z 2019 *Nature* **566** 480
[62] Maia G. Vergniory, Benjamin J. Wieder, Luis Elcoro, Stuart S.P. Parkin, Claudia Felser, B. Andrei Bernevig, and Regnault N arXiv preprint (2021). arXiv:2105.09954.
[63] Disalvo F J and Waszczak J V 1980 *Phys. Rev. B* **22** 4241
[64] Disalvo F J and Waszczak J V 1980 *J. Phys. Chem. Solids* **41** 1311
[65] Boubeche M, Yu J, Chushan L, Huichao W, Zeng L Y, He Y, Wang X P, Su W Z, Wang M, Yao D X, Wang Z, Jun, and Luo H X 2021 *Chin. Phys. Lett.* **38** 037401
[66] Yang J J, Choi Y J, Oh Y S, Hogan A, Horibe Y, Kim K, Min B I, and Cheong S W 2012 *Phys. Rev. Lett.* **108** 116402
[67] Boldrin D and Wills A S 2012 *Adv.Cond. Matter. Phys.* **2012** 615295
[68] Shi M Z, Yu F H, Yang Y, Meng F B, Lei B, Luo Y, Sun Z, He J F, Wang R, Wu T, Wang Z Y, Xiang Z J, Ying J J, and Chen X H 2021 arXiv:2110.09782

Table 1. Lattice parameters, atomic coordinates, and isotropic thermal factors $B_{iso}$ for $V_3Sb_2$ samples from powder XRD data at 295K.

|  | S1 (780 °C) | S2 (830 °C) |
|---|---|---|
| Space group, $Z$ | $R\text{-}3m$, $Z = 6$ | |
| $a$ (Å) | 5.5440(1) | 5.5545(1) |
| $c$ (Å) | 20.3529(4) | 20.3374(6) |
| $V$ (Å$^3$) | 541.76 (2) | 543.40 (2) |
| V_$x$ | 0.49470(15) | 0.49480(25) |
| V_$y$ | 0.50530(11) | 0.50520(25) |
| V_$z$ | 0.11018(11) | 0.10734(15) |
| $B_{iso}$_V (Å$^2$) | 2.65(5) | 2.32(3) |
| Sb1_$z$ | 0.33189(4) | 0.33281(5) |
| $B_{iso}$_Sb1 (Å$^2$) | 2.72(3) | 2.41(3) |
| Sb2_$z$ | 0.11101(06) | 0.10908(08) |
| $B_{iso}$_Sb2 (Å$^2$) | 3.49(4) | 2.76(5) |
| $R_p$ (%) | 2.66 | 2.98 |
| $R_{exp}$ (%) | 2.48 | 2.26 |
| $\chi^2$ | 2.28 | 3.78 |
| $R_{Bragg}$ (%) | 4.35 | 4.20 |

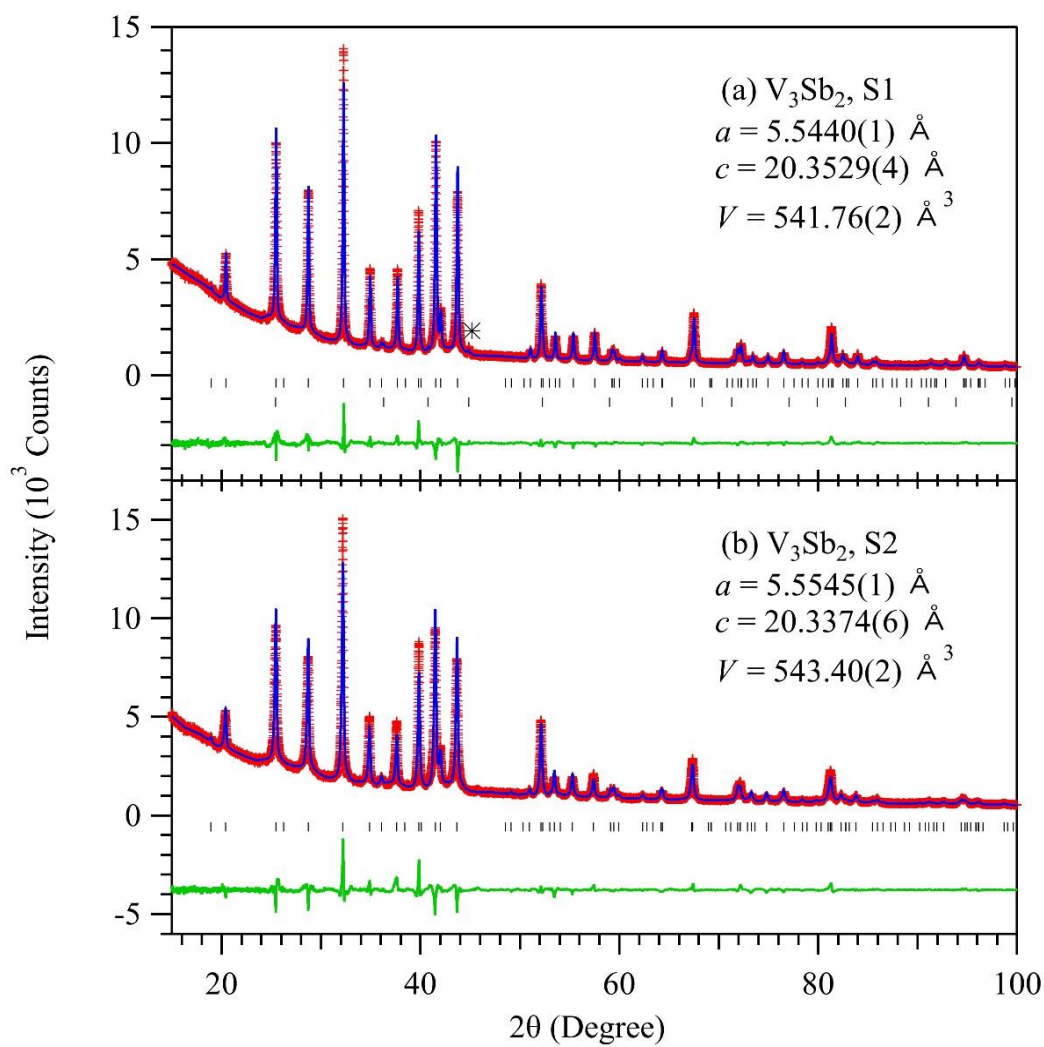

**Figure 1**. Observed (cross), calculated (solid line), and difference (bottom line) XRD profiles of the polycrystalline $V_3Sb_2$ samples: (a) S1 and (b) S2 after Rietveld refinements. Bragg positions of the main phase $V_3Sb_2$ and the impurity phase $V_3Sb$ are indicated by the two rows of tick marks in (a), while only the position of $V_3Sb_2$ is shown in (b). The main peak of the $V_3Sb$ impurity phase is marked by an asterisk in (a). The obtained lattice parameters for both samples are also given in the figure.

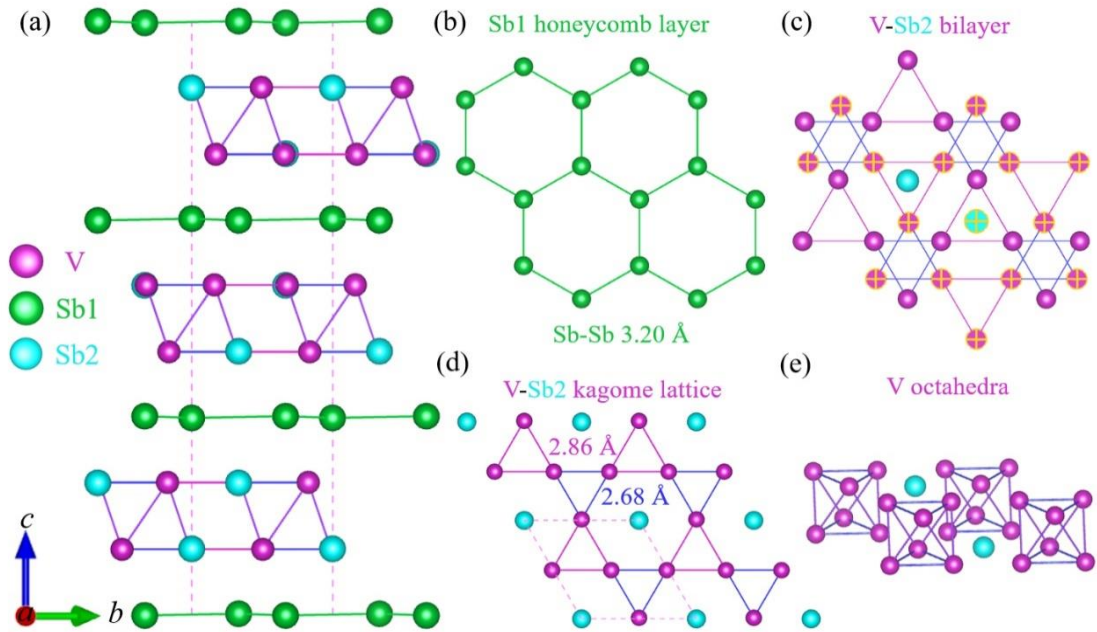

**Figure 2**. (a) A schematic view of the layered structure of $V_3Sb_2$ (side view) stacked by Sb1 and V-Sb2 layers along the *c*-axis. The unit cell is enclosed by the dashed lines. (b) The Sb1 atoms form a honeycomb sublattice below and above each V-Sb2 bilayer with Sb1-Sb1 bond length of 3.20 Å. (c) The structure of the V-Sb2 layer. The V atoms form bilayer kagome lattice and the Sb2 atoms sit in the centers of V hexagons in each V kagome layer. (d) Single V-Sb2 kagome layer made of two kinds of equilateral triangles with the side length of 2.8602 Å and 2.6839 Å, shown in purple and blue, respectively. (e) The V triangles with smaller length (blue) in two layers form the $V_6$ octahedra. The bond-length values are taken from the refined results of the S1 sample

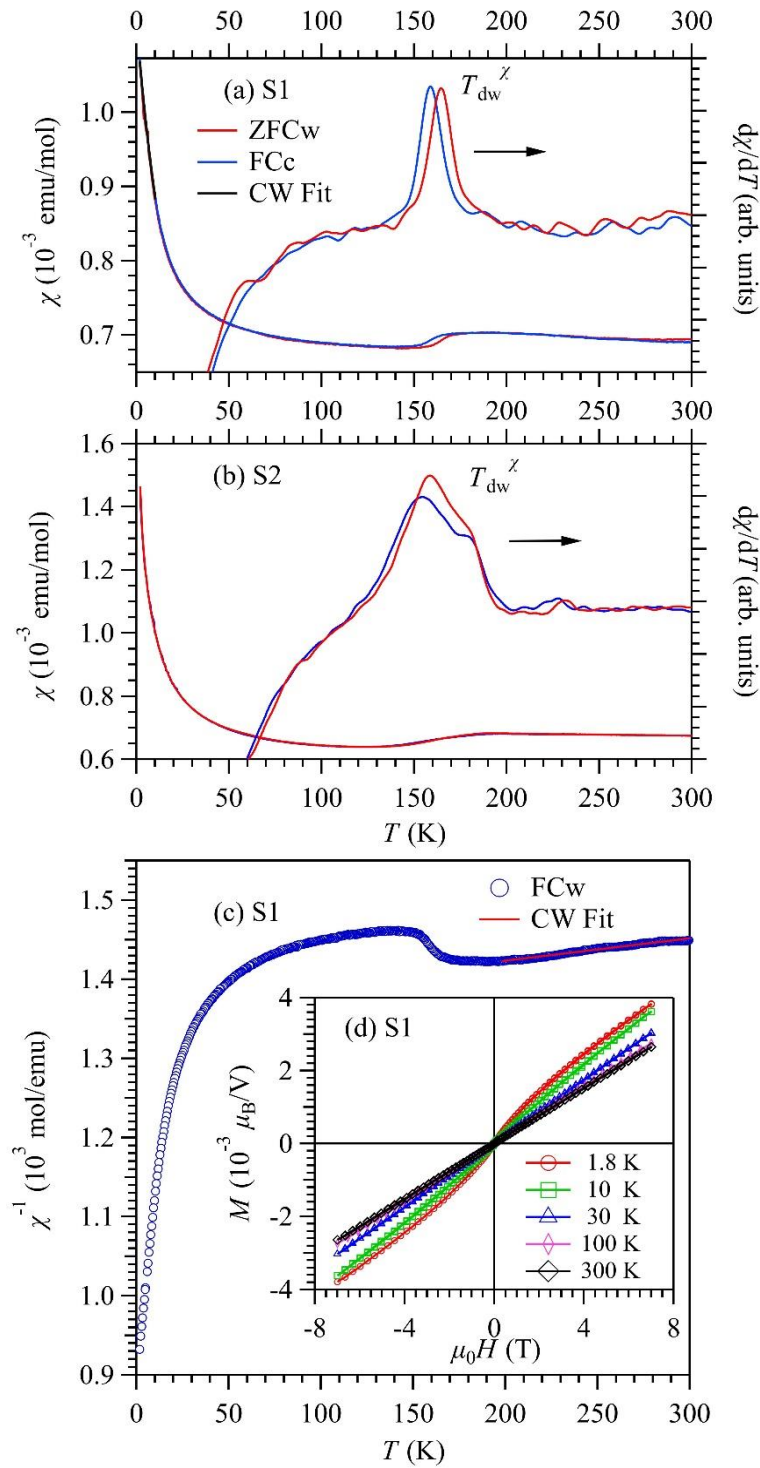

**Figure 3**. Temperature dependence of the dc magnetic susceptibility $\chi(T)$ and its derivative $d\chi/dT$ for $V_3Sb_2$ samples: (a) S1 and (b) S2, measured in the ZFCw and FCc modes under an external magnetic field of 1 T. The transition temperature $T_{dw}^{\chi}$ was defined as the peak of the $d\chi/dT$. (c) Temperature dependence of the inverse susceptibility $\chi^{-1}(T)$ for sample S1. The Curie-Weiss (CW) fitting curves are shown by the solid lines in (a) and (c). (d) The isothermal magnetization $M(H)$ curves for sample S1 measured between +7 and −7 T at various temperatures.

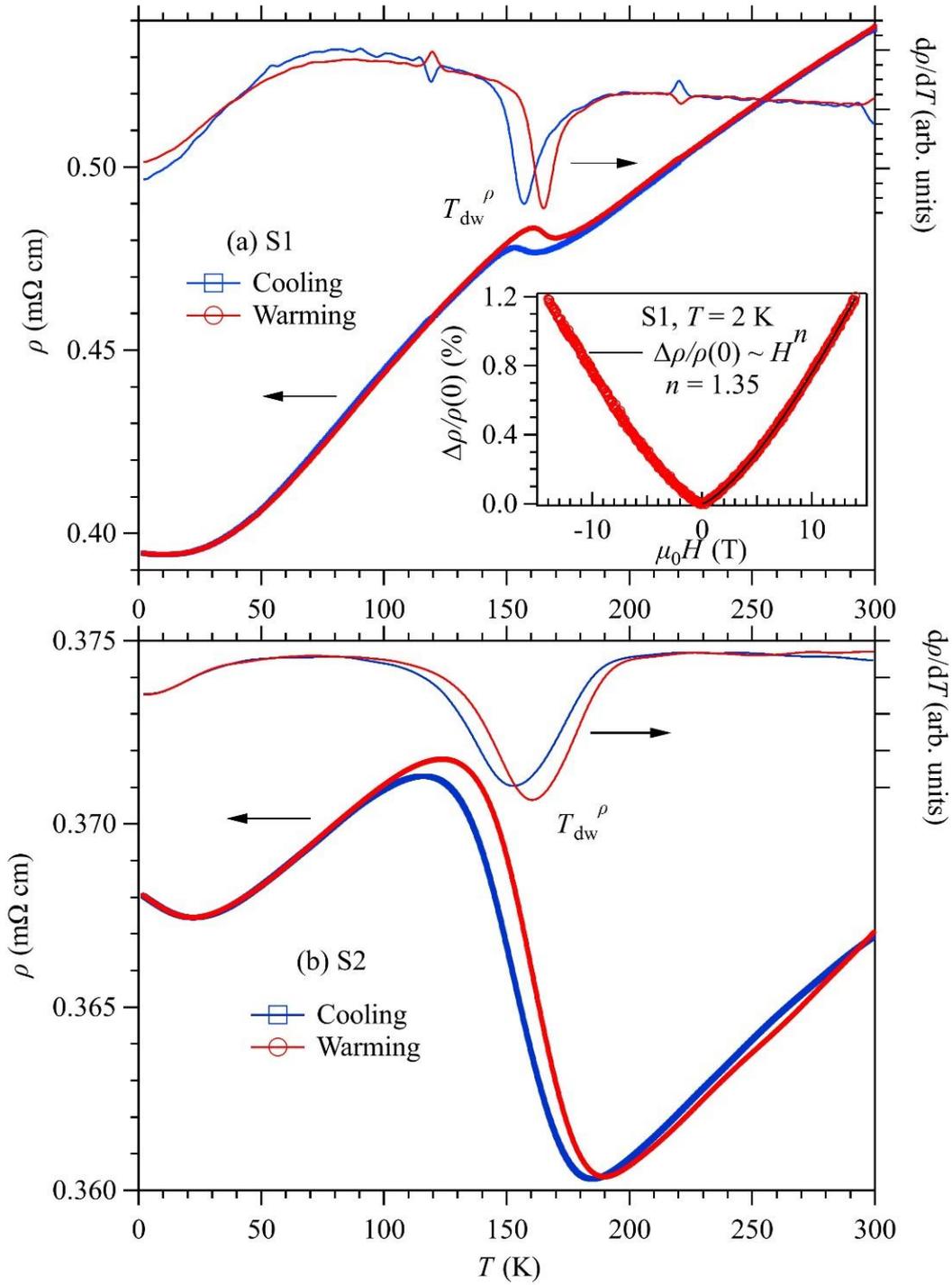

**Figure 4**. Temperature dependence of the resistivity $\rho(T)$ and its derivative $d\rho/dT$ for the $V_3Sb_2$ samples: (a) S1, (b) S2, measured in a thermal cycle under 0 T. The transition temperature $T_{dw}^{\rho}$ was determined from the dip of $d\rho/dT$. Inset of (a) shows the magnetoresistance of S1 at 2 K under fields up to 14 T.

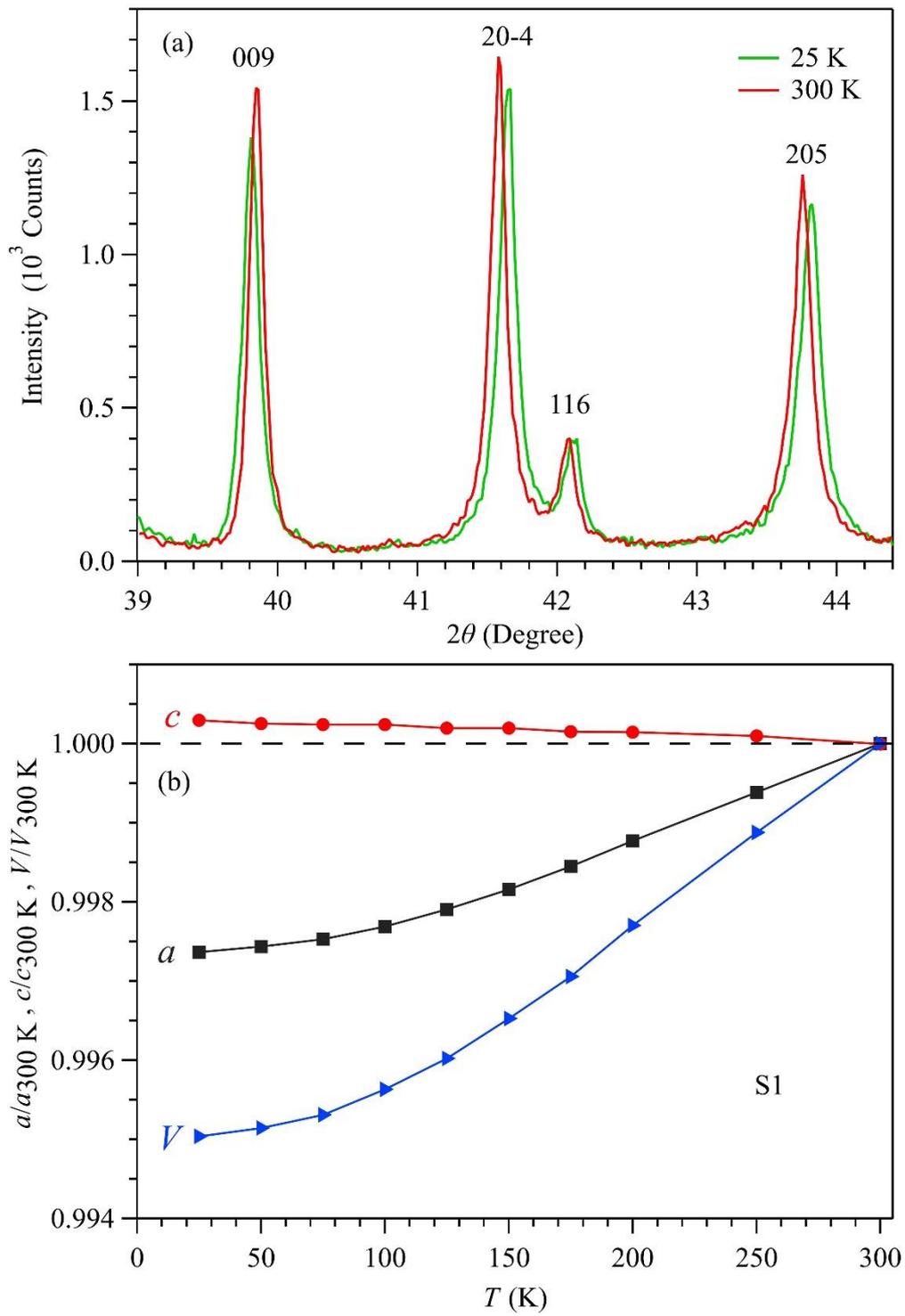

**Figure 5.** (a) A portion of the XRD patterns measured at 25 and 300 K. (b) Temperature dependences of the normalized lattice parameters of $V_3Sb_2$ sample S1.

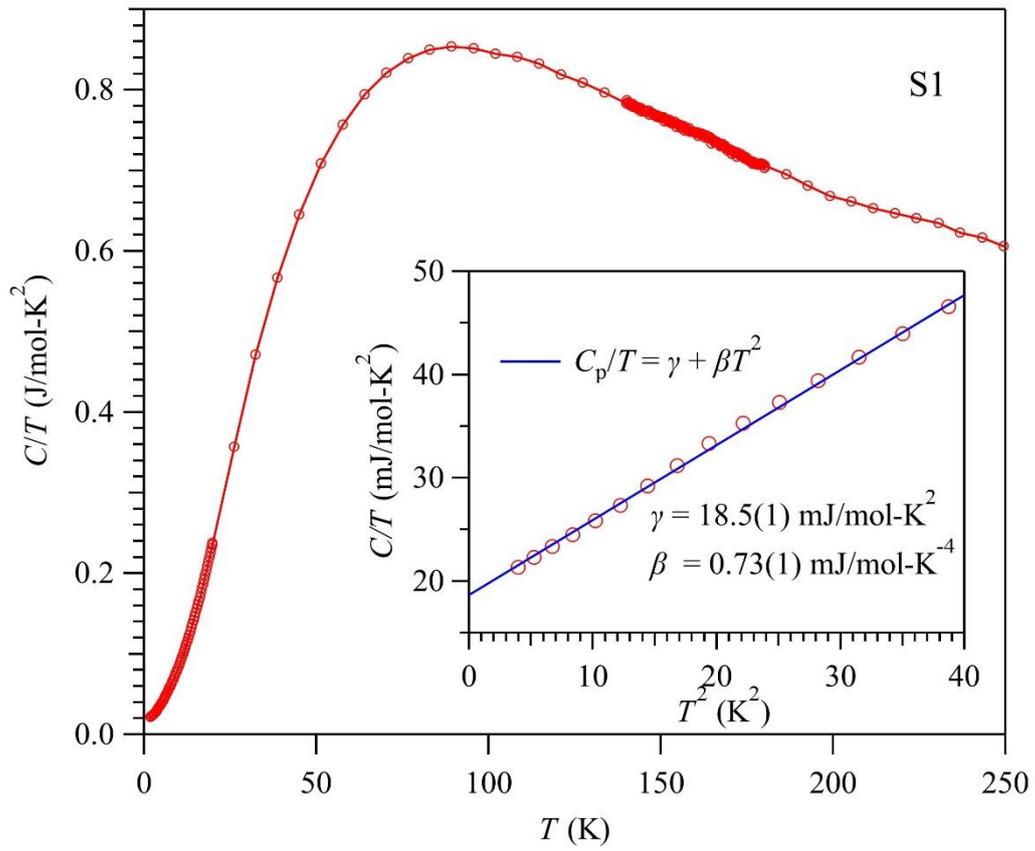

**Figure 6**. $C/T$ vs $T$ for $V_3Sb_2$ sample S1 in the wide temperature range from 2 to 250 K under zero field. Inset shows the $C/T$ vs $T^2$ at the low-temperature range, which can be described by a sum of electronic and phonon contributions.

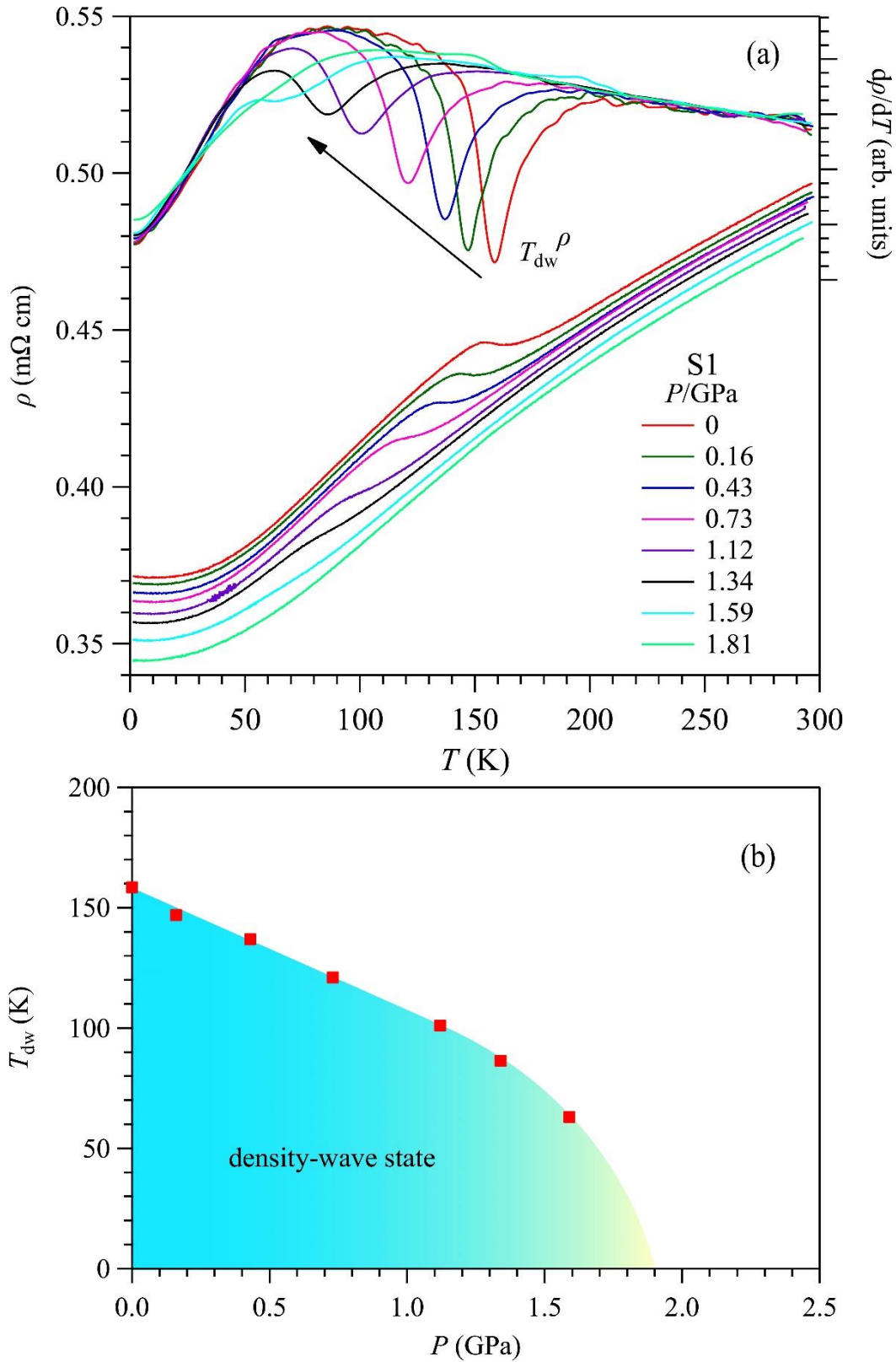

**Figure 7**. (a) Temperature dependence of the resistivity $\rho(T)$ and its derivative $d\rho/dT$ for the $V_3Sb_2$ sample S1 under various pressures up to 1.81 GPa measured with a piston-cylinder cell. (b) Pressure dependence of the density-wave-like transition temperature $T_{dw}$.

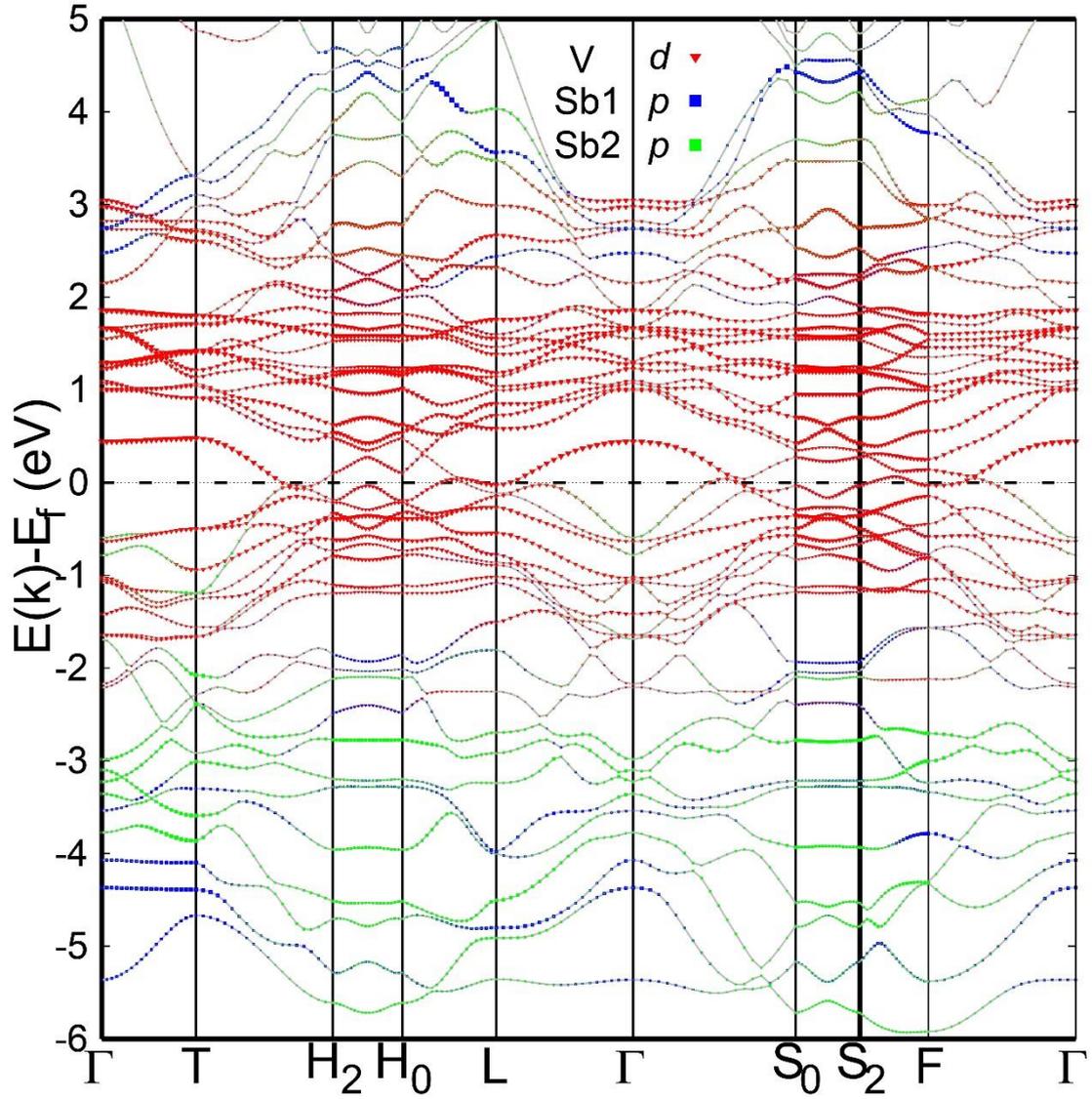

**Figure 8**. The projected band structures obtained from DFT calculations for $V_3Sb_2$. The orbital characters of bands are represented by different colors and the projected weights are represented by the sizes.